# AFFORDANCES OF DIGITAL AND BLOCKCHAIN-BASED COMMUNITY CURRENCIES: THE CASE OF SARAFU NETWORK IN KENYA

Patricia Marcella Evite[1]

[1] *University of Paris Cité, Graduate School of Sustainability and Transitions 5 rue Thomas Mann, 75013 Paris, patriciaevite@gmail.com*

**ABSTRACT:**

Community currencies (CCs) have been adopting innovative systems to overcome implementational hurdles from issuing paper currencies. Using a qualitative approach, this paper examined this digital transition of Sarafu Network in Kenya and its predecessor CCs as a case study. From the original vouchers launched in 2010, the foundation Grassroots Economics introduced a digital interface in 2016 that operates on a feature phone, and then integrated blockchain technology starting in 2018, undergoing several migrations before becoming settling on its current iteration called Community Asset Vouchers on the Celo blockchain since 2023. Using affordances from human-computer interaction, the research shows that digitalization and blockchain improved the facilitation of economic activities of the local communities, both their typical market transactions as well as traditional reciprocal labour exchanges, by offering more functionalities compared to the analog version of Sarafu. The unique contributions of blockchain include enabling automation of holding tax calculations and linking the vouchers to the mainstream monetary system via stablecoins facilitated by a series of smart contracts also known as the liquidity pool. The study also finds that there is an inherent trade-off between blockchain benefits and user interface complexity. Hence, balancing innovation and community needs remains a challenge.

## 1. INTRODUCTION

The term "community currency" (CC) might imply inherent inclusivity, given its focus on populations often lacking access to traditional financial systems. In this paper, I am more interested in CCs as financial products (Telelbasic, 2017) that have physical and technical characteristics that influence their accessibility to underserved people as a means of payment. As money became increasingly virtual, the focus on CC's tangible form often obscures the labour or "moneywork" that underpins the successful functioning of these systems (Perry and Ferreira, 2018).

With the advent of Bitcoin, CCs have outsourced accounting and other functions to the blockchain technology, giving rise to "solidarity cryptocommunities" (Diniz et al., 2020). Some examples include those based in the United States like FairCoin (Kasliwal, 2019), Kolektivo City3, CityCoins or Ibiza Token (Rong & Dam, 2023), Auroracoin in Iceland (Allen, Berg & Lane, 2023) or the GIZ-supported payment platform called 'Our Village' in Cameroon (Wu et al., 2023). Blockchain operates as an ever-expanding, collaboratively maintained, immutable, and resilient historical ledger. Conceptual work on blockchain's potential to enhance CC transparency and governance is more extensive (see Scott, 2016; Takagi et al., 2017; Kalmi, 2018; Siqueira et al., 2020) than real-world studies, the latter relying heavily on prototypes, simulations, or small-scale pilot projects (Orzi et al., 2021; Hasna et al., 2022; Szemerédi and Tatay, 2020). Thus, the actual operational dynamics of blockchain-based community currencies remain underexplored.

This research aims to bridge this empirical gap through a qualitative case study on the Sarafu (currency in Swahili) Network in Kenya. Established in 2010 by the non-profit Grassroots Economics Foundation (GE), Sarafu has evolved over a decade of designs involving paper vouchers, digital, a combination of both, and at present, blockchain-based tokens since 2018. The present iteration which was launched in 2023, called Community Asset Vouchers (CAV), was finalized. This system leverages the Celo blockchain, designed for mobile and low-cost payments, and operates through two user interfaces: 1) an Unstructured Supplementary Service Data (USSD)[1] that allows transactions on basic mobile phones without internet access and; 2) non-custodial wallets, i.e., accounts that give experienced users full control of their private keys, often the one of the main purported benefits of going on the blockchain. Sarafu's present hybrid design reflects GE's attempt to balance accessibility with user autonomy, an issue central to financial inclusion in low-connectivity environments.

To assess the impact of these technological advancements, the study employs the concept of affordance theory from the field of human-computer interaction, which explain the potential actions of a technology for its users. We use this as proxy for benefits of a CAV format. Accordingly, the central questions guiding this research are:

1. What is the context in which digital platforms and the blockchain are used in the Sarafu Network?
2. What are the affordances offered by each CC format from paper, digital and blockchain-based?

The study is novel in two ways. First, it approaches the research question from a new perspective. The study's use of affordance theory exhausts the improvements by looking at all possible actions of the users from the digital and blockchain-based infrastructure. Second, this paper focuses on the use of CCs for Rotating Labour Associations (ROLA), a traditional, non-monetized form of reciprocal group work. While Ba et al., (2022) looked at transaction patterns by group-issued CAVs, it did not account for the reciprocity involved in their social agreements. Since the practice of ROLAs has dwindled with the rise of wage labour and individualization of societies (Wilson, 2001; Shiraishi, 2006), its continued use in rural communities (Karanth, 2002; Manosalvas et al., 2021) makes its connection to a digital technology nearly a nondocumented phenomenon. This research, therefore, specifically contributes insights into how the digital transition can support the sustenance of these traditional resource coordination practices.

The remainder of this paper proceeds as follows. Section 2 establishes the empirical context of the Sarafu Network and the Mweria reciprocal labour system. Section 3 presents the theoretical foundations of the moneywork lifecycle and affordance theory. Section 4 details the qualitative research methodology, scope, and limitations. Section 5 presents the findings from the fieldwork, and Section 6 concludes.

## 2. CASE STUDY CONTEXT

### *2.1. Technological Evolution of the Sarafu Network*

Up to this point, the vouchers have been considered as CCs, but the foundation and the stakeholders identify them as vouchers, more akin to gift cards, that are redeemable for the issuer's products and services. Specifically, the vouchers are considered by GE as "formalized representation of commitments" (Ruddick, 2023). Resistance to labels reflect the tension associated with legal tender. In addition, money has a speculative aspect, thus distorting its notion (Stepnicka et al., 2020) as unit of account and medium of exchange (Dini, 2014). To respect these preferences, the programs of the foundation will be generally referred to as vouchers, but comparisons to general literature will still include that of CCs and other forms of money where necessary.

Phase 1: Paper Vouchers (2010-2016)

The initiative began with Eco-Pesa in 2010 backed by donor funds which evolved into the well-known business surplus-backed Bangla-Pesa in 2013 and other paper-based vouchers in other communities. By 2017, they were consolidated into Sarafu Credit with a 1:1 exchange rate with other users (Ussher et al., 2021). While successful in facilitating local trade, this analog system was beset by significant logistical and financial hurdles. These included high printing and security costs (e.g., watermarks), physical deterioration, and, most critically, the logistical burden of executing the holding tax for the members through manual renewal stamps. Users also forgot the paper bills at home (Barinaga and Zapata Campos, 2023)

Phase 2: Centralized Digital (2016-2018)

To solve the problems of paper, GE introduced Sarafu-Credit, a digital system built on a familiar USSD (text-based) front-end via feature phones, requiring no internet connection or smartphone along with a centralized Structured Query Language (SQL)[2] database back-end. This shift eliminated printing costs, but the back-end was costly to maintain as well and thus, did not last long.

Phase 3: Decentralized Blockchain (2018-Present)

To address the issues of cost, centralization, and governance, GE began migrating its back-end to blockchain technology. This migration was not a single event but a multi-step evolution:

- 3a. Bancor (2018) (Ussher et al., 2021): The first attempt involved a what is called the Proof-of-Authority (PoA) blockchain, where only approved operators are allowed to validate new transactions and add new blocks. It also used the Bancor protocol to manage exchange rates among multiple vouchers (e.g., Bangla, Miyani, etc.) using a bonding curve mechanism, allowing algorithmic control over redemption rates, similar to an automated currency board.
- 3b. xDai (2020): In response, GE moved the system to the xDai sidechain (now Gnosis Chain) to reduce transaction cost (Ba et al., 2022). This also meant dropping the bonding curve, and consolidating back to a single token called Sarafu (Ussher et al., 2021). This period also was notable for the Red Cross-backed 100 Sarafu Credits distributed during the pandemic and the introduction of a monthly 2% fee as a holding tax to encourage spending (Ussher et al., 2021)
- 3c. Kitabu (2022-2023): After xDai shifted to a Proof-of-Stake (POS) system, which imposed gas fees to prevent overload in the system, the foundation ran its own ledger using a POA scheme called Kitabu, meaning ledger in Swahili (Ba et al., 2022). The goal was to maintain free transactions for its users.
- 3d. Celo (2023-Present): The final migration was to the Celo blockchain, a platform chosen for its explicit mobile-first design and humanitarian focus. Celo allows GE, acting as a network validator, to earn rewards in the form of cUSD, a stablecoin pegged to the US dollar. A validator runs nodes that confirm transactions and propose new blocks, earning rewards for securing the network. GE leverages this reward stream to fund its free-transaction model for users.

### *2.2. Mechanics of ROLA systems*

While businesses and individuals use Sarafu, the study's unit of analysis is the chama, a group of 10–30 members who pool resources, extend credit. Although originally designed as a complementary CC for small businesses with surplus goods and services (Bendel, Slater, and Ruddick, 2015) or tool for humanitarian assistance (see Ussher et al., 2021; Ba et al., 2022) Sarafu and its predecessor CCs have been inadvertently used as instruments for reciprocal labour (Ruddick, 2023), a non-monetized exchange of group work within a community for the benefit of its members (Gibson, 2020). Groups that practice these are also called rotating labour associations (ROLA). In this paper, we use ROLA interchangeably with Mweria, the local term by selected indigenous tribes in Kenyan coastal communities.

This system is distinct from a "gift economy" which is based on altruistic giving without expectation of an equivalent return (Lemmergaard and Muhr, 2011). Instead, ROLA is categorized as a "balanced exchange" (Sahlins, 1972, as cited in Gunasinghe, 1976), a "tit-for-tat" system that demands "strict reciprocity" (Hames, 1987). Macfarlan, Remiker & Quinlan (2012) conceptualized competitive altruism instead as the psychology behind ROLAs, where individuals display generosity not only for others but also to secure cooperative relationships or status within their group. It also goes beyond a single family unit (i.e., that of familial labour (Guillet, 1980)) and often involves broader social networks or community arrangements. The motivation for ROLAs include "returns to teamwork" especially in farm tasks (Gilligan and Lopez, 2004), prompted by abundant land but limited capital and supplies (Wilson, 2001).

Although the work is done collectively, reciprocity in ROLAs is individualized, where each instance of labour time exchanged between two team members represents two sides of the same bilateral transaction (Gilligan and Lopez, 2004). Because of this, the system requires the formal tracking of balances as the group membership increase. This study draws on the definitions of balance from Hames (1987):

1. General Balance: The system-wide goal, where the absolute difference between what a member gives and what they receive deviates from zero. The objective is to return this balance to zero. The CAVs then function as a way to track how much everyone owe each other, consistent with the definition of CC as "mutual credit" (Bendell, Slater and Ruddick, 2015).
2. Specific Balance: The individualized, matrix-based tracking of who owes how much to whom. If a member fails to participate, they owe a specific labour credit to that host.

Barinaga (2020) considers chamas as a governance institution of their own monetary systems. For example, repeated non-observance of the mutual obligation can suspend one' membership.

## 3. THEORETICAL FRAMEWORK

### *3.1. Moneywork Lifecycle*

Collavecchia (2005) conceptualized moneywork as the often-invisible domestic labour involved in managing family finances from earning, budgeting, saving, to spending. Perry and Ferreira (2018) then applied it in Human-Computer Interaction (HCI) setting by identifying the interactional work around the use of money in financial transactions involving preparatory activities, technical constraints, practices of use, and social interactions, thus adding the "lifecycle aspect" (see Table 1). Their empirical analysis with now defunct Bristol Pound, a mixed media CC supporting mobile payments, revealed that what appears as a simple moment of exchange actually involves extensive "articulation work" across time, requiring users to coordinate social, technical, and financial elements. Because it examines the hidden and non-digital aspects of a transaction, the moneywork lifecycle has been particularly useful in studying digital payment systems with vulnerable and marginalized populations (see Muralidhar (2019) and Barros Pena et al. (2021)).

*Table 1. Parts of the Moneywork Lifecycle*

| Stage | Focus (Activity) | Key Labour Involved |
|---|---|---|
| **Pre-transaction** | Preparatory Activities | Spending decisions, ensuring fund readiness (appropriate accounts/forms), and maintaining payment mechanisms (devices/cash). |
| **At-transaction** | Immediate Interactive Exchange | Assessing the payment environment, coordinating terms, managing technical requirements, executing value transfer, and confirming completion. |
| **Post-transaction** | Follow-up and Completion | Security maintenance, financial housekeeping (tracking/records), and social activities (relationship maintenance). |

*Source: Adapted from Perry and Ferreira (2018)*

Since this study deals with a CC and not a legal tender, the transactional processes are distinct from those of cash or digital money. A CC's value is often tied to goods and services, and its exchanges involve a host of social interactions absent in general economic activity. Therefore, to analyze this unique flow, this study adapts the three-stages. We note that this paper utilizes its chronological division purely as an analytical structure. This provides a necessary, discrete framework to understand how technology supports these exchanges and at which specific stages that technology is involved.

### *3.2. Affordances Theory*

In Human-Computer Interaction (HCI), the affordance theory explores the relationship between goal-oriented actors and an IT artefact (Hazra & Priyo, 2020). Adopted from ecological psychology, the concept was introduced to HCI by Donald Norman, expanding on Gibson's original idea of affordances as "properties of the environment relative to an animal" that indicate potential actions (Jenkins et al., 2008). Affordances theory were introduced to HCI from ecological psychology to enhance the understandability and usability of artifacts, particularly their user interface.

In finance, affordances theory has been used as a discovery framework for stablecoins (Szabo et al., 2022) and broader blockchain-based financial systems (Kim, 2023), respectively. More directly related to this study's empirical approach is Hazra and Priyo (202)'s study of bKash users in Bangladesh, providing a methodological precedent. We follow the former's adoption of "potential affordances" stance, i.e., affordances as potentials emerging from technology-user relationships vis-a-vis affordances directly embedded in the IT artefact's design.

Potential affordances are latent action possibilities not yet realized, while actualized affordances emerge through users' active perception and engagement. According to Lehrig (2017), potential transforms to actual use through configuration that is either delegated, guided, or autonomous. Thus, affordances are not fixed but dynamically shaped by users' expectations, technological capabilities and contextual conditions (Shin, 2022).

## 4. EMPIRICAL DESIGN

### *4.1. Data Collection*

A two-month immersion period between March and May, 2024 (Table 2) was conducted in Kilifi, Kenya, primarily engaging with two chamas (Figure 1). Data collection involved one (1) participatory workshop with the Miyani chama, who met the criteria of having experience with both paper and digital versions of CAVs. The Miyani chama were asked to reflect on the differences between those two experiences, with the aid of local translators as necessary. At the time of the field work, there were 72 out of 74 chamas who practiced ROLA at least once. Convenience sampling was utilized due to limited public transportation and high costs associated with expanding the sample. For all other field activities, the main chama observed was Kiriba. When referring to the issued CAV of the chamas, which usually follow their group name as well, I use capitalized letters, i.e., MIYANI is the CAV of the Miyani chama.

*Table 2. Timeline of Activities*

| **Activity** | **Mar 1-15** | **Mar 16-31** | **Apr 1-15** | **Apr 16-30** | **May 1-7** | **Location(s)** |
|---|---|---|---|---|---|---|
| Immersion and Observation | | | | | | Miyani, Kilifi |
| Participatory Workshop | | | | | | |
| Staff Interviews | | | | | | GE Headquarters (HQ) and remote |
| Supplementary Data Gathering | | | | | | GE HQ and Kinango Kwale |
| Mweria & Jubilee Events | | | | | | Kiriba, Kilifi |

*Source: Authors*

Each chama has chairperson, who also owns the wallet address or Subscriber Identification Module (SIM) card where the community fund goes to; a treasurer if the group is big enough to necessitate one; a secretary and a champion or field officer whose job is to liaison with GE. Usually, champions are members who own or at least know how to operate a feature phone to facilitate voucher transfers and help in trouble shooting technical problems and other issues in the chama. The champion and/or chairperson keeps logs of the balances at each meeting.

Additionally, five (5) semi-structured interviews were conducted with technical staff (developers) tagged as [D], both remote and in-person, where possible, and one (1) focus group discussion was held with the community development or field officers tagged as [G] (See Figure 2). Developers were mostly asked about the processes from both front and back end technologies of Sarafu Network. Supplementary materials from the Moneyless Society collective and Insight Share (2024) who conducted documentary film shootings in the same period. They helped corroborate and expand on reported Mweria practices and are accordingly tagged as [S].

*Figure 1. Demographic overview of Kiriba and Miyani chama participants*

*Source: Authors*

Kiriba members are older and more ethnically mixed (see Panel D), while Miyani members are younger, and culturally homogeneous. Both groups have predominantly female members because males are engaged in wage labour.

*Figure 2. Demographic overview of GE Staff Members*

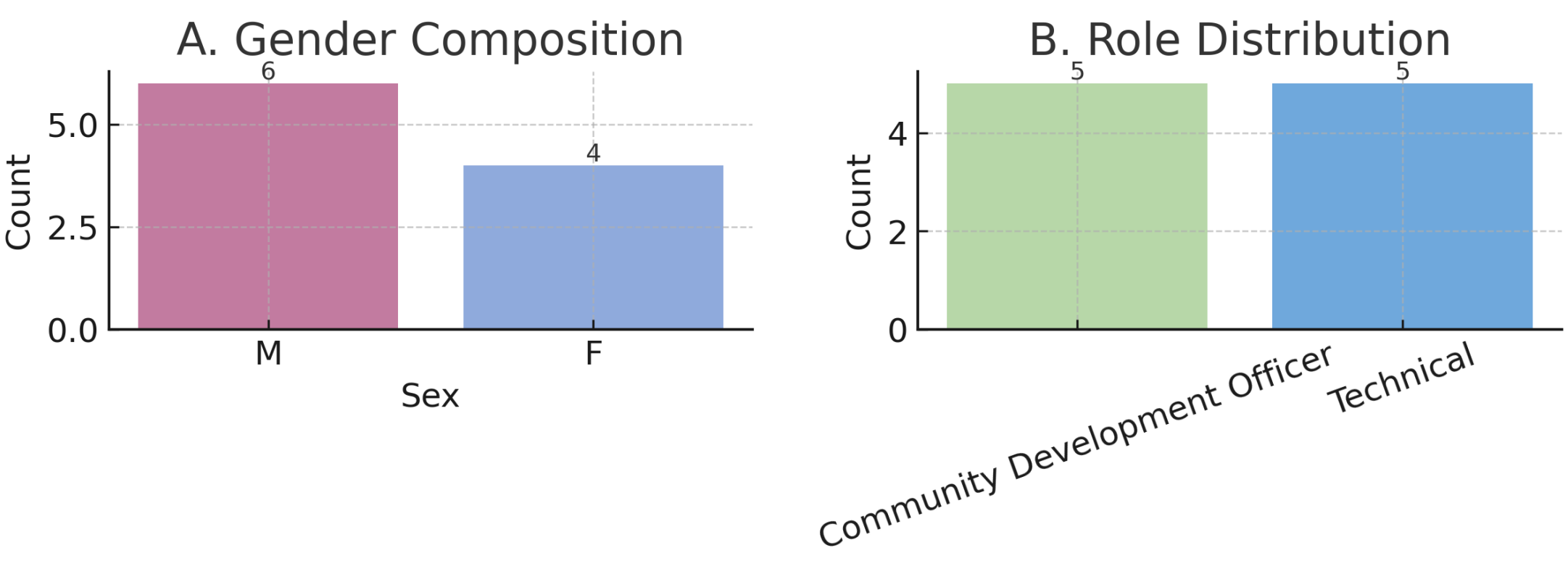

*Source: Authors*

Age, roles and total years active were redacted because the size of the organization risks revealing the participants' identity. Technical staff had at least 2 years involvement with GE and 3 years for the community development officers. The former were a mix of past and present employees and were necessary to cover the whole stage of technological evolution of CAVs.

### *4.2. Ethics Statement*

Explicit informed consent was gathered and signed by the participants prior to their participation in the data collection. Where necessary, it was translated to their local language in written and oral form to accommodate also illiterate members of the community. It is also important to acknowledge potential biases that may have influenced the research process since the author resided in the accommodation provided by the foundation. This close proximity to the organization may have inadvertently influenced the researcher's perceptions and interactions with participants. Efforts were made to mitigate this bias by maintaining a reflexive approach throughout the research process through supervisions with more senior scholars in the field as well as turning to literature and secondary data.

### *4.3. Scope and Limitations*

All iterations of GE's CCs — from Eco-Pesa, Bangla-Pesa, Sarafu-Credit, to Sarafu—are treated here as part of the same system, collectively CAVs. While their names, governance, and technologies have evolved, the analysis focuses on their formats (paper, digital, and blockchain-based) rather than viewing them as entirely distinct currencies. Accordingly, references to the blockchain[3] are used in a general sense to denote the distributed ledger infrastructure supporting Sarafu's digital systems. Specific blockchain versions (e.g., Bancor, xDAI, Celo) are mentioned only when clearly identified by participants. This approach constitutes a minor limitation, as some platform-specific or design-level variations may not have been fully captured. Because this is an empirically derived process, it may be the case that some activities are not observed during the immersion and therefore, the results from may not paint a complete picture of all the work required and all use cases for CAVs.

Crucially, the study's use of affordance theory is strictly descriptive: it reveals the actions and possibilities that the technology enables, such as the fungibility of labour or automated demurrage. The findings do not constitute an ultimate, prescriptive evaluation of whether that enablement is universally beneficial for community development, as this requires consideration of the accompanying social and technological trade-offs.

## 5. FINDINGS

### *5.1. The Digital and Blockchain in a CAV Payment Process*

The entire CAV transaction process is analyzed using the Moneywork Lifecycle framework, which structures the flow into three chronological stages to pinpoint where the technology intervenes (Figure 3).

*Figure 3. Moneywork Lifecycle of a CAV*

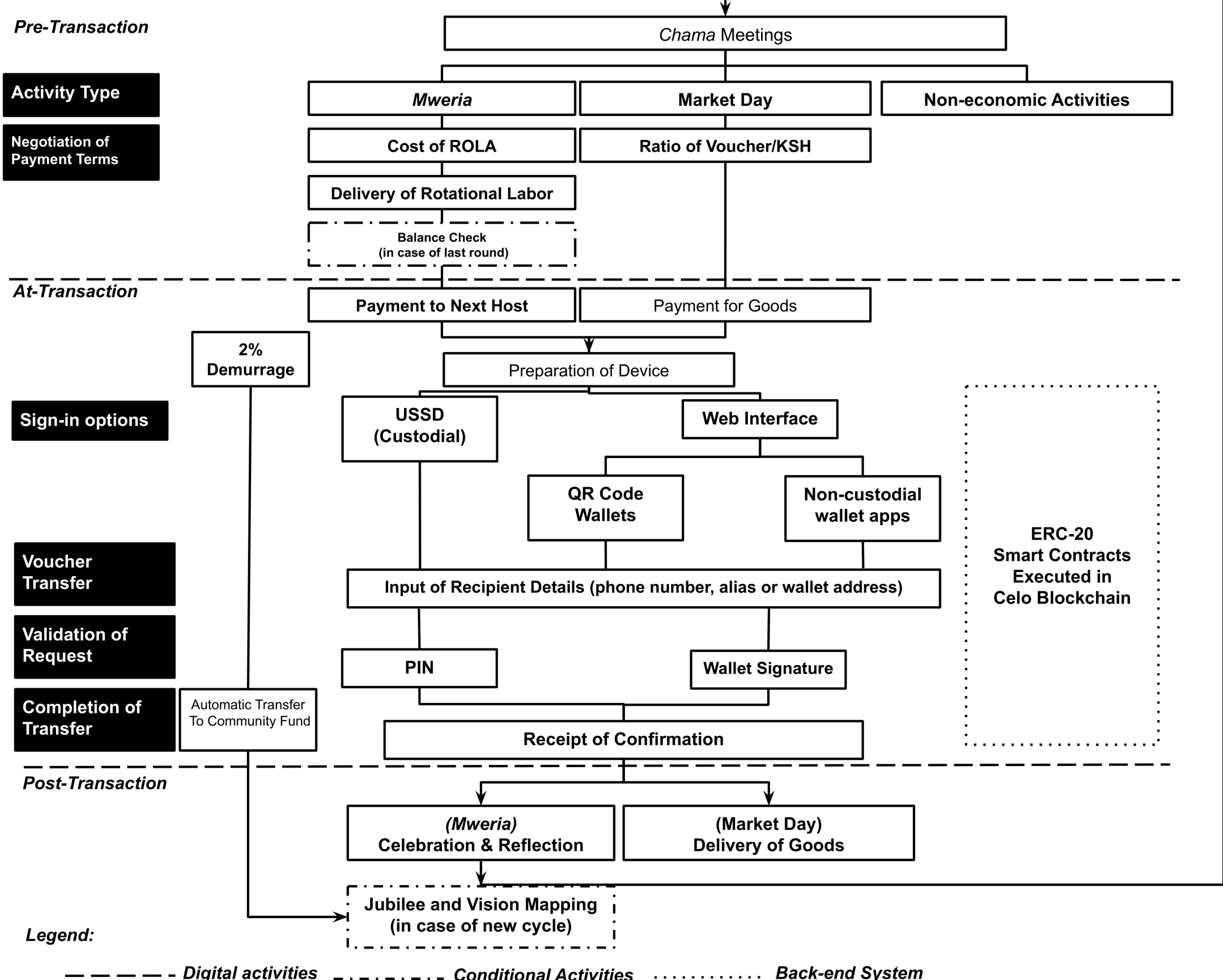

*Source: Author's illustration, adapted from Perry and Ferreira (2018). Note: Dashed lines denote the digital activities, which include the flow of transactions and digital sign-in options. Dash-dotted lines represent conditional activities that only occur at the conclusion of a Mweria cycle. Dotted lines delineate back-end system of the latest iteration, which is the ERC-20 smart contracts.*

### *5.1.1. Pre-transaction*

The Pre-Transaction stage consists of all preparatory activities necessary before an exchange can be executed. These activities are anchored in the chama meetings, which are regular gatherings where members engage in Mweria, market day exchanges (buying and selling products), and non-economic activities including fellowship or advice giving by the elders to various familial and societal problems encountered. We focus on the first two for this study.

- Mweria Negotiation: The group collectively decides the order of the rotational labour, prioritizing those with urgent needs. To ensure the reciprocal labour contribution is equal, members use time as a unit of account. As with any economic transaction, price must be determined; payment for ROLA is negotiated by the community and typically falls between 10–20 vouchers as observed with the Kiriba chama.
- Market Day Negotiation: For buying and selling products, the negotiation involves the seller suggesting the Ratio of KES/CAV, which usually does not fall below 50:50 (in favour of KES).
- (Conditional) Balance Preparation: If it is approaching the last round, a balance preparation ensues with the objective of achieving a general balance of zero before the next cycle starts. Members with a positive balance are required to actively purchase goods or services from other members until their account hits zero debit. Conversely, those holding a negative balance are required to commit to bringing products to the next meeting or arrange for later delivery. In some instances, the chama may opt to use the community fund to clear a negative balance in the spirit of solidarity.

### 5.1.2. *At-Transaction*

The At-Transaction stage covers the immediate interactive work of executing the exchange. The transaction is used for two purposes: to express gratitude for reciprocal labour or settle debt obligations, and to pay for goods provided by other members.

- Access and Initiation: Users initiate the payment by selecting one of three sign-in options (Figure 4). The options include the low-tech USSD (Custodial) system (panel A), a web-based interface accessed via QR Code Wallets (panel B), or connecting to non-custodial wallet apps (panel C).
- Voucher Transfer and Validation: For transfers, the user inputs the Recipient Details (phone number, alias or wallet address). The request is validated by the user's Personal Identification Number (PIN) for USSD or a wallet signature for web apps, a digital equivalent of signing a check, proving the transaction request came from the wallet owner. The latter is a cryptographic confirmation, generated using your wallet's private key, that proves you are the owner of the funds and approve the transfer.

*Figure 4. Three Sign-in Options for Sarafu Network*

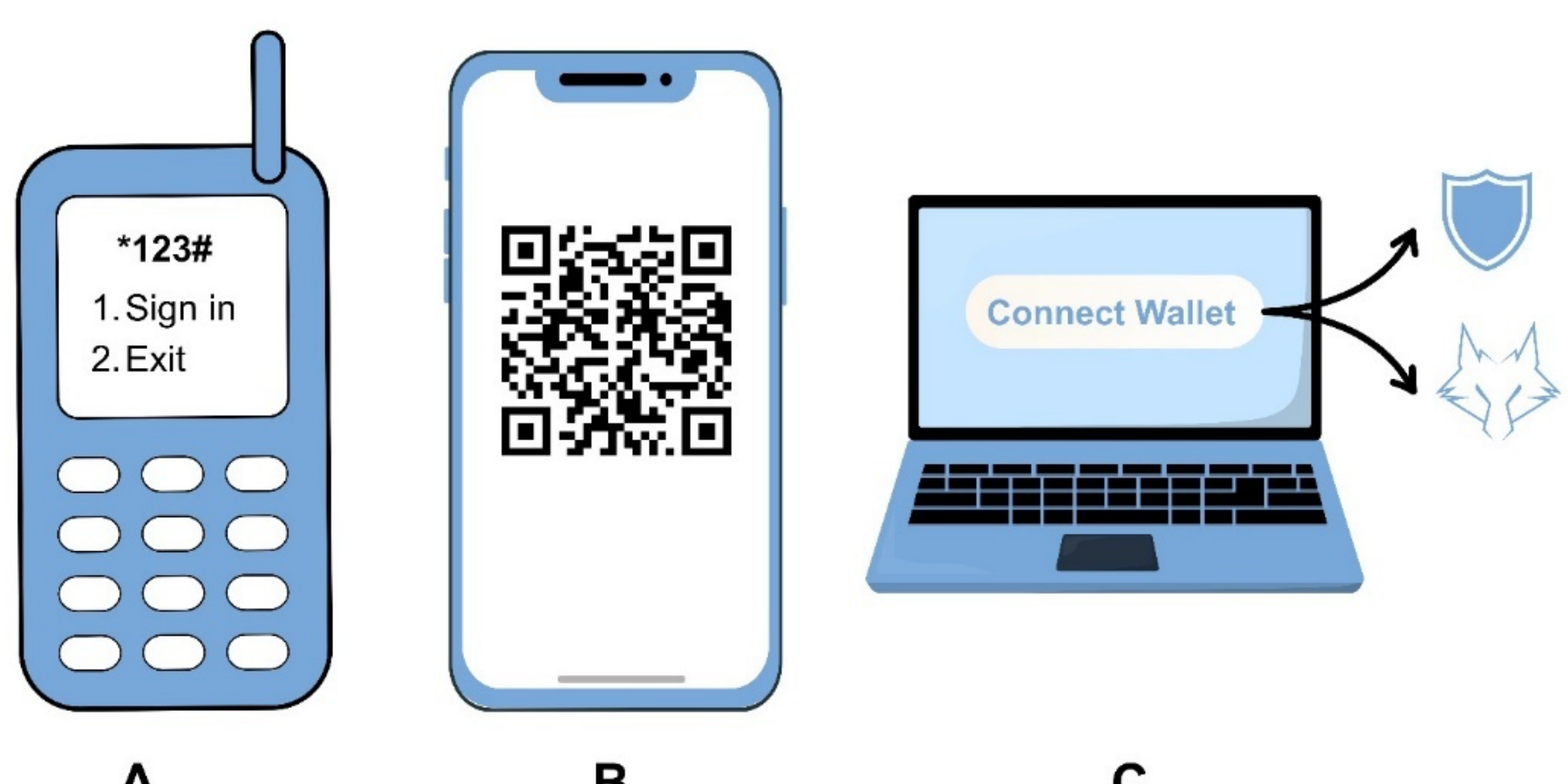

*Source: Visualization by Alexandra Beatrice Brion from author's input*

- Back-End and Automation: In the current back-end system, all transfers are executed via ERC-20 Smart Contracts Executed in Celo Blockchain. The smart contract checks parameters like ensuring sufficient balance. Crucially, the smart contracts automate demurrage calculations, monitoring time using blockchain timestamps to deduct the fee and automatically transfer it to the Community Fund.
- The payment concludes with the Completion of Transfer, resulting in a Receipt of Confirmation.

### 5.1.3. *Post-Transaction*

When the group reaches the final round of the Mweria cycle, members hold a Jubilee, a celebration marking the completion of all reciprocal obligations. The Jubilee functions as both a social and economic reset: balances are reconciled, any outstanding debts are settled, and the community collectively evaluates what worked and what must change for the next cycle. This process of reflection is sometimes complemented by Vision Mapping, a practice led by Grassroots Economics facilitators to project future collective goals, such as organizing new training, expanding membership, or preparing for the next agricultural season. The social fabric with which the Mweria-practicing chamas operate in distinguishes them from those who do not because market transactions have a finality. In contrast, ROLAs are marked in the figure by arrow connecting it to the beginning of the transaction lifecycle once again, representing continuous involvement.

### 5.2. Affordances

#### 5.2.1. Shared Affordances: Value Expression Flexibility

The introduction of vouchers enabled communities to express their gratitude for the help in a fungible manner, offering similar advantages to cash:

- "When we got the voucher, we...help each other with the voucher instead of cooking for the group members.[S]"
- "...because we used to cook food from your home for the group members but we changed to use voucher and it used to go around. When it reaches tomorrow (day of Mweria), they do send the voucher to me even if they won't manage to come for Mweria. Once they are done with the work of the day, I will send them the voucher.[S]"

Even in humanitarian contexts, cash-based assistance is increasingly favoured over physical goods due to this flexibility, providing choices to recipients on where they spend it (Heaslip et al., 2016). But of course, this does not guarantee that it is spent towards the initial goal of a humanitarian program. By converting from meals to vouchers, the benefit can also be considered only at the individual level if spent for their own or own family's goods and services. But because of the social agreements made at the chama level, the voucher is claimable for the previous host's labour, thereby allowing other members to use their own skills in access for others.

The other possible effect of vouchers is that it replaced the meal for convenience and thereby also decreased the time that would have been for breaking bread together. In the limited time of the period observed, this was not the case since the Mweria constituted other non-economic activities before each ROLA was performed. The value expression flexibility affordance is thus aligned with what Blanc (2017) posits that primitive forms of money were designed for social exchanges and contribute to their circulation, challenging the common belief that reciprocity is incompatible with monetary transactions (see Lietaer & Meulenaere (2003)'s study on the Balinese banyar in Indonesia where they find that money is not substitutable for voluntary cooperative labour).

#### 5.2.2. Shared Affordances Digital Interface and Blockchain

##### 5.2.2.1. Accessibility

In the context of the digital transition for the Sarafu Network, achieving accessibility is a conditional affordance, reliant on mediating social and infrastructural factors. For example, in contrast to Kiriba with 100% phone ownership, Miyani only has 15 of 27 members with phones. The remaining 12 (44%) only had SIM cards, giving rise to an autonomous configuration to avail the affordance — a "communal phone", where phone owners lend devices and switch SIM cards to facilitate post-Mweria voucher transfers.

However, another constraint to affordance is the inability of illiterate members to perform transfers. The mitigation is the deployment of "champions", typically younger and more digitally literature members who act as troubleshooters and/or perform the transfers on behalf of others. GE provides personalized training and continuous support, noting that "there's a period of hand-holding whereby we still visit them, attend their Mwerias to ensure that they are smooth on their own [G3]." This can be seen as a mix of autonomous (appointment of champions) and delegated (organizational support) configuration towards accessibility.

While paper vouchers offer better accessibility (requiring only comprehension of numbers and no phone) the digital system benefits from existing familiarity with Kenya's M-Pesa USSD platform as observed also by Barinaga and Zapata Campos (2023), which is more accessible than QR codes or web interfaces. Staff and developers alike share this opinion: "...just like the way they use like M-Pesa so it's the same with [that]. They use the USSD so many are familiar [with the] steps to follow in order to send like that money.[G2]""....it's better to use a...custodial system, using a very simple interface like USSD because they are used to that.[D3]"

*5.2.2.2. Transaction Visibility*

The shift to digital vouchers improved transparency allows for better monitoring and analysis of transactions. Prior to the introduction of paper vouchers, communities within the chama system relied on informal ledgers to track liabilities amongst members. These ledgers essentially functioned as records of IOUs (a phonetic acronym for I owe you). The vouchers unintendedly replaced this system. On the other hand, the initial voucher allotment acted as the businesses' tracking system. A surplus of vouchers indicated credit owed to the community, while a deficit signified a debt requiring repayment:

"Each and every person were given 400 [vouchers][4] so it was easy to track. If you have more than 400, you know that [if you have an extra] 100, the 100 is not yours. If we have less than 400, you know that there is a hundred or 50 somewhere...that it is yours...so you have to track it yourself and see where the [voucher] went. [M]"

Transitioning to digital contributed to reporting and data visualization abilities of the foundation. Even in the USSD system, there is a log of transactions, enabling circulation studies like Mattson, Criscione and Takes (2023), that are otherwise untraceable during the era of paper vouchers. However, with the blockchain, all transfers are recorded, and it enabled the foundation to construct dashboards. Network visualizations became possible, deepening the understanding of chama interactions and leading to the discovery of key relationships, including members acting as brokers, and the identification of routine transactions that revealed the existence of Mweria reciprocal labour practices (Ruddick, 2023).

However, this potential affordance of the blockchain was actualized thanks to the high technical programming skills by the foundation. This is required due serialization of data, a technical process of converting complex digital data structures and objects such as transaction details (e.g., sender, receiver, amount) into a standardized, linear stream of bytes. As such, transactions need to be decoded and processed to extract useful information from the Sarafu network like sender, receiver, and amount. In a traditional or SQL database system, the stored information is often the same as the one you collected in raw form, but this is not the case for blockchain. A programmer notes: "...a human cannot go into the blockchain and decode and say, "X" did send this much to that person. You need to pull that information out and process it and then "humanize" it.[D4]"

*5.2.3.1. Spatial and Temporal Mobility*

Similar to Hazra & Priyo (2020), we found that users perceived spatial and temporal mobility affordances, by bypassing the constraints when using paper vouchers. Transactions can occur for market purchases and Mweria reciprocal labour or debt settlements without physically handing over vouchers, like the shift to digital money (Tomita, 2022; Calhoun et al., 2019).

- "You could use it (digital vouchers) when you are not around, maybe you're in Nairobi, and your people are in Miyani, and they are out of cash, you can send them the voucher and then they can buy food using the voucher. [With] paper voucher, you can only use it [while you are] in Miyani. [M]"
- "With the paper voucher, you use[d] to walk to the place where you want to make a purchase from...walk a distance...but then with the digital voucher, you don't need to walk because you can just get the number of that person and just dial and send there the amount for the commodity that you want to purchase...that was her biggest change she has seen. [M]"

By offering access via the Web Interface (using QR Code Wallets or Non-custodial wallet apps), the system is no longer reliant on the local telecommunication network coverage, a limitation inherent to USSD. This wider reach allows for remote issuance and transfer for any user with a Celo-compatible wallet, facilitating external use cases and supporting members who travel outside the core local service area.

*5.2.3.2. User-facing Security*

Instead of relying on a familiarity-based authentication of the members like with Bangla-Pesa (Pusateri, 2013), digital vouchers performed the verification of who is authorized to use them. While it was manageable for a small chama, it may not be the case as memberships scale: "During the paper voucher, a small number was registered...so it was easy to track who owns the paper voucher...anyone having the paper voucher who wants to buy goods and services and is not the [authorized] person, it was easy to identify...they know either...[that it is his] or her parent is the one registered...specifically supposed to use that voucher. [M]"

Another benefit is that counterfeit with the digital vouchers is difficult because faking a USSD session takes a high level of expertise. The most one can do is "sending an SMS to somebody to pretend that it was that (a CAV transaction) [D5]". However, the same level of simplicity that enhanced usability also introduced a new vulnerability. Limited literacy skills led elders to verbally communicate their PIN numbers in the presence of others when seeking assistance from literate members to execute transactions: "If you don't know how to use USSD, and you give it to someone who knows how to use USSD, it was easy for that person to steal [for] you. [sic] [M]". Cordelia Rose and Yap (2020) reported similar issues in consumer vulnerability where limited literacy skills hinder protection and lead to challenges in interpreting labels and making decisions. To counter account compromise, Sarafu incorporated a built-in "social" recovery system, allowing users to appoint guardians to reset their PINs, a mechanism that worked effectively within chamas due to strong social bonds and regular meetings.

The subsequent shift to a blockchain back-end introduced a trade-off between accessibility and security affordances. Given the foundation's mandate for inclusion, the choice to offer an unencrypted wallet vis-a-vis one with a password (Figure 4) actualizes the former, which is good news for the digitally nascent users, but not for the security latter affordance in that there is no added layer of protection like the PIN in the USSD method. Moreover, adopting a non-custodial model meant that lost private keys or passwords are unrecoverable.

*Figure 5. Web interface for creation of paper wallet in Sarafu Network*

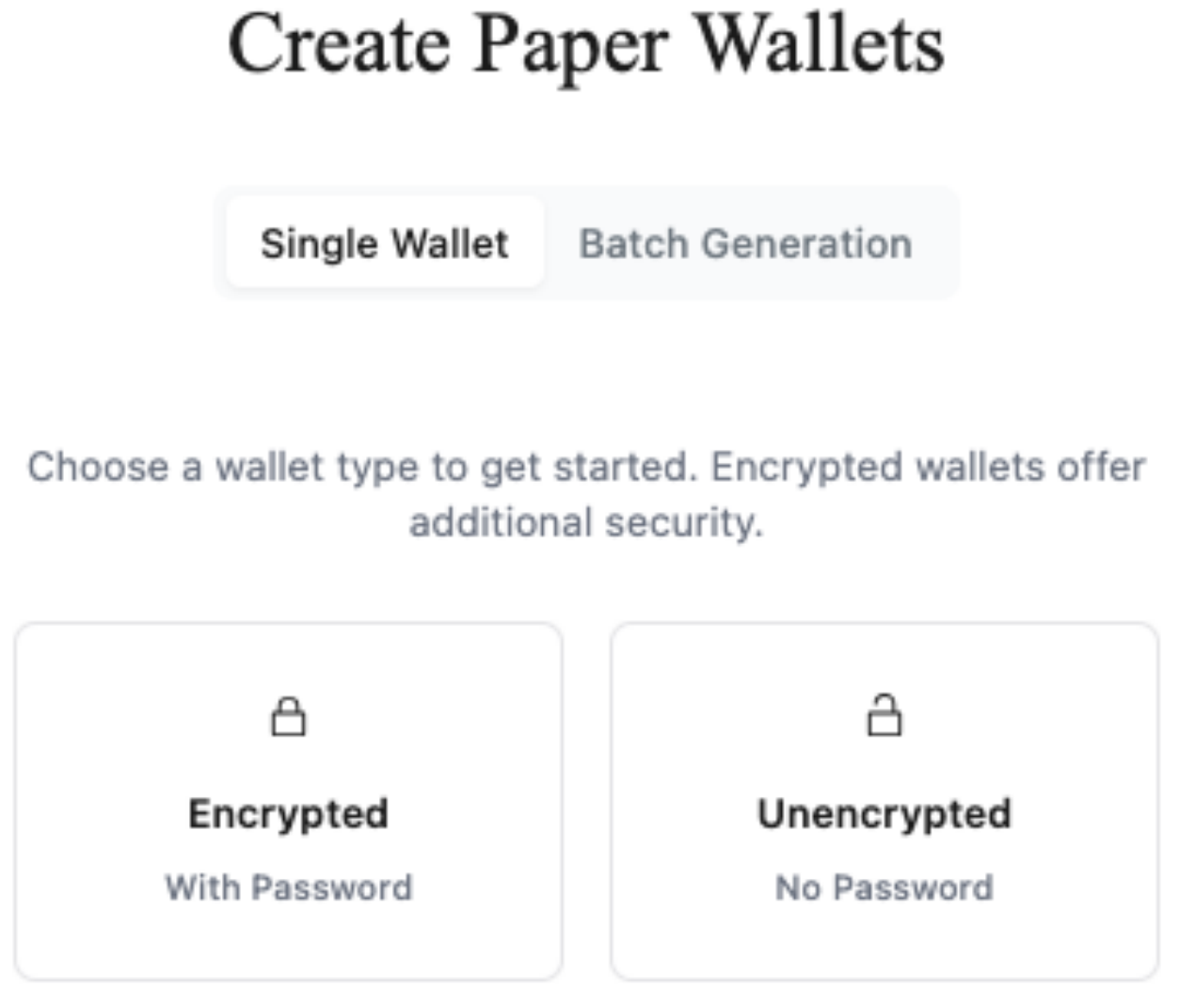

*Source: https://sarafu.network taken 2 November 2025*

### *5.4. Blockchain Affordances*

#### *5.4.1. Organizational Accountability*

Before running the transactions on the Celo blockchain, Sarafu transactions were temporarily on a centralized server operated by the foundation. While such a higher degree of administrative control does not necessarily equate to insecurity, a closed, in-house structure posed more a single-point governance failure. On the pre-Celo version where GE ran its own nodes, there was also a higher chance of history rewriting and denial of service attacks if those operators were compromised or lacked redundancy. While internal misuse was not reported, a respondent comments on the theoretical risk, "...if we were like four people in the organization, running all the nodes, then obviously we would also have the power to, if we have the know-how to, also the power to counterfeit. [D5]"

By transferring to a broader network like Celo with more stakeholders, there is also an inclusion of a third non-interested party that increases the computational power that renders counterfeiting not only costly, but practically impossible. Furthermore, data resilience is actualized because data is replicated across multiple nodes: "...we can't just lose it (data) overnight as in a traditional database [where] you need to take care of disaster management. On a blockchain, as long as you have...people [with] same copy...you can rebuild all the transactional data and just continue from where you left off...[D4]"

#### *5.4.2. Programmable Functionalities*

The defining affordance of the blockchain back-end is Programmable Functionality: the ability to customize and control the voucher system's rules through code (smart contracts). This capability birthed two key affordances.

##### *5.4.2.1. Automated Demurrage Calculations*

The shift to digital vouchers eliminated the administrative burden associated with currency maintenance. Demurrage (a voucher holding tax) was a manual process involving members physically renewing their membership with GE staff, which was prone to error and discouragement due to the hassle. A participant noted the benefit of the digital transition: "The paper voucher had some like expiration, so it needed to be stamped on so that at least to be continued used. But then this one, the new one, which is in digital, you only have that voucher in your phone and you don't need to renew it or to be stamped on, to be stamped, yeah. And you can still use it, yes. [M]"

These automated programs now handle demurrage calculations based on predefined rules. When a voucher is created, the demurrage rate (e.g., 2% daily), expiry date (e.g., 30 days), and the community fund address (where fees are sent) are set within the contract. The contract continuously tracks time using blockchain timestamps and automatically transfers the calculated demurrage to chairperson's account. This replaces the need for physical bookkeeping and treasury with traceable records.

##### *5.4.2.2. Commitment Pool*

Commitment Pools is the organizational term used by the foundation to describe a smart contract mechanism that functions as a liquidity pool. A liquidity pool is an automated market-making smart contract that holds balances of multiple tokens contract that holds the balances of multiple tokens and enforces rules around its deposit and withdrawals. For vouchers in such a pool, users can perform swaps, e.g., one can exchange one voucher for another, at a predefined exchange rate. Once the users give permission for the contract to spend their vouchers, it deducts the specific amount required to withdraw the equivalent of another token, automatically deposited back to the user's wallet. The goal of this feature in the context of chamas is to make it possible for an individual to partake in economic activities of another chama by putting their own voucher in exchange for the new one. This provides an infrastructure for the scenario of Sarafu users with multiple memberships acting as brokers (Ruddick, 2023).

Moreover, GE has introduced a seed option for donors to invest financing into the liquidity pool in the form of Celo USD (cUSD). cUSD is a stablecoin native to the Celo blockchain and designed to maintain a stable value pegged to the US dollar. As a validator in the Celo blockchain, GE received cUSD as rewards for participating in the network. They deposit them back into these pools for chamas to access or use it to waive the gas fees associated with the transaction as part of their mission. It can be seen as a gateway to national currency because users, once withdrawn a cUSD, can convert them into KES in Celo-compatible trading apps.

In the example in Figure 5, a donor or organization seeds the pool with 1,000 cUSD, joining other CAVs namely RIBA, SRF and MIYANI. The user then deposits 10,000 RIBAs in exchange for 75 cUSD. Because this voucher pool system was launched in May 2024, further research is needed to fully assess its potential benefits. While the current sources for stablecoins now are just solely the foundation, initial observations still offer promising insights. First, chama members of Kiriba were able to exchange their vouchers for the cUSD initially seeded in one of the ROLA pools newly established. After exchanging the cUSD with KES, they bought seeds and distributed them to the chama equally. This demonstrates the potential of voucher pools to facilitate transactions outside the chamas and the ROLA system through stablecoin conversion.

*Figure 6. Swap Process in a Commitment Pool*

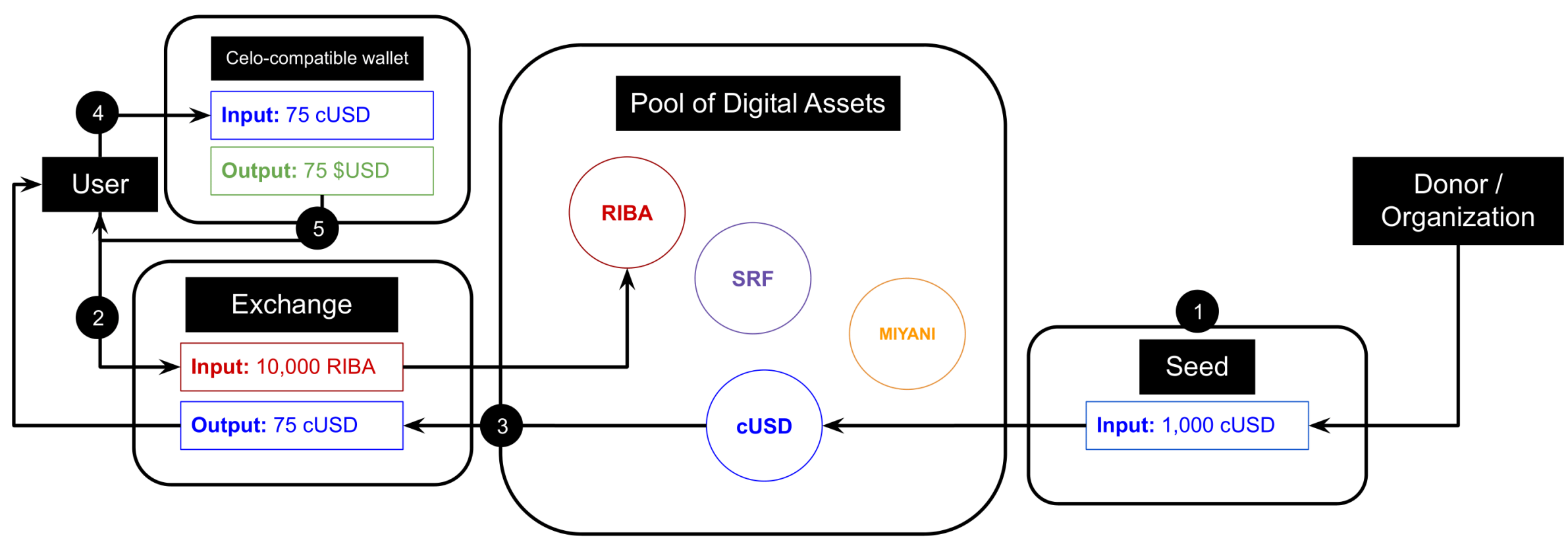

*Source: Author's illustration, adapted from Uniswap (n.d.)*

## 6. SUMMARY AND CONCLUSIONS

The technological evolution of Sarafu's Community Asset Vouchers (CAVs) fundamentally actualized the affordance of value expression flexibility, concurrently demonstrating that social cohesion can successfully embed and sustain a digital currency. This paper explored how the digital and blockchain-based evolution of the Sarafu Network reshaped the affordances of community asset vouchers (CAVs) and their implications for local exchange systems in Kenya through a qualitative approach. Using Human-Computer Interaction (HCI) lenses, we identified a total of seven (7) affordances that are shared and distinct among the three technological formats of the CAVs: paper, centralized digital, and blockchain-based. All CAV types actualized the affordance of value expression flexibility, replacing the previous need to prepare meals or offer in-kind payments for reciprocal labour exchanges (Mweria), and instead providing workers with a value they could utilize as they saw fit for their economic needs. Although this rendered the festive part of ROLA unnecessary, the very nature of Mweria kept intact the group's cooperative relations. Moreover, the social cohesion of the chama, which enabled the effective security of the digital system via social recovery, fundamentally challenges the notion that money makes relations colder and individualistic. This supports ideas that incorporation of indigenous knowledge are not irrelevant but rather lead to contributions to financial systems (Bray and Els, 2007) because it allows for the creation of more appropriate financial products and services (Johnson et al, 2014).

Digital systems introduced key affordances like enhanced mobility and transaction visibility, yet these benefits were contingent upon social and technical workarounds to overcome local infrastructure and literacy constraints. Both the digital front-end and blockchain access (web sign-in and backend) provided accessibility, transaction visibility and monitoring, user-facing security, and spatial and temporal mobility. These digital systems enhanced transactions by offering the capability to settle Mweria obligations 'anytime, anywhere,' overcoming the spatial constraints of paper vouchers. Nonetheless, this shift was not without challenges, including phone requirements, network issues, and limited literacy. These constraints were configured or mitigated by the social commitments of the chamas and support from actors, including the foundation and the champions, who introduced training and workarounds, such as phone-owning members lending their devices or printing unencrypted QR codes on paper, eliminating the necessity for a second scanning device.

These digital systems also presented new security vulnerabilities, from verbally shared PINs to unrecoverable non-custodial keys, where the mitigation factor was primarily social trust, not technology. While the centralized USSD system enhanced security through PIN authentication and a pre-validation step displaying transaction details, low literacy sometimes inhibited members' ability to protect themselves, leading to instances where elders verbally communicated their PIN numbers in front of others. Furthermore, the introduction of the non-custodial blockchain model meant that lost private keys or passwords are unrecoverable, which is a frequent issue reported to customer support. Moreover, prioritizing user inclusion led to the adoption of unencrypted QR code wallets, creating a security vulnerability where unauthorized access to the code effectively bypasses password protection. A crucial mitigation factor specific to this community context, however, is the reliance on established social trust within the chama, allowing members to negotiate and re-issue vouchers in cases of lost access.

The blockchain architecture uniquely introduced crucial business and programmable affordances, including organizational accountability via immutable records and enhanced utility through automated processes. The blockchain uniquely introduced new organizational accountability as a business affordance through publicly verifiable and immutable transaction records, enhancing back-end security by distributing risk across multiple nodes. The programmable functionality affordance enabled the automation of demurrage calculations, transferring fees directly to the community fund address and eliminating the cumbersome manual renewal process. This functionality also enabled the establishment of Commitment Pools (liquidity pools) that linked community currencies to stablecoins (cUSD). This bridge to the mainstream monetary system provides users with flexible access to fiat currency for necessities (like buying seeds or paying fees).

The findings also highlight that decentralization comes with more constraints to actualizing the affordances by the technology (Figure 6). The affordance analysis confirms that a technological enablement is not uniformly "good." To maximize these affordances, however, takes a highly technical village of engineers and developers, raising the persistent concern of financial inclusion. Barinaga (2020) also notes that the "autopilot money governance" offered by crypto infrastructure at the expense of "communal democracy" erosion. On the other hand, democratizing the code makes it independent of the foundation's leadership in line with the concept of "participatory design of token economies" (Avanzo et al., 2023). The hybrid configuration of Sarafu, combining USSD-based custodial systems and non-custodial wallets, thus represents a pragmatic compromise between innovation and inclusion. The core finding is ultimately that the adoption of digitalization and blockchain technology improved chama transactions by providing more affordances than paper-based vouchers.

While NGOs like Grassroots Economics provide essential technological innovation, the greatest gaps that inhibit the affordances (i.e., educational and digital literacy) lies within the duty of public institutions. Multiple studies found that government support can help vulnerable groups to user financial technologies (Okello Candiya Bongomin et al., 2025; Mishra et al., 2024). Otherwise, the digital divide would perpetuate existing inequalities. Buchanan (2014) suggests that governments can support alternative economies in indigenous communities by recognizing and making customary economic activities more viable through policy and activism, challenging state and market dominance, and using existing surveys and studies to inform support for community food economies, that we can relate to Mweria-practicing chamas. It is also found to be essential in cultural resilience and cultural preservation (Mistry et al., 2023).

*Figure 7. Trade-off between affordance constraints and decentralization*

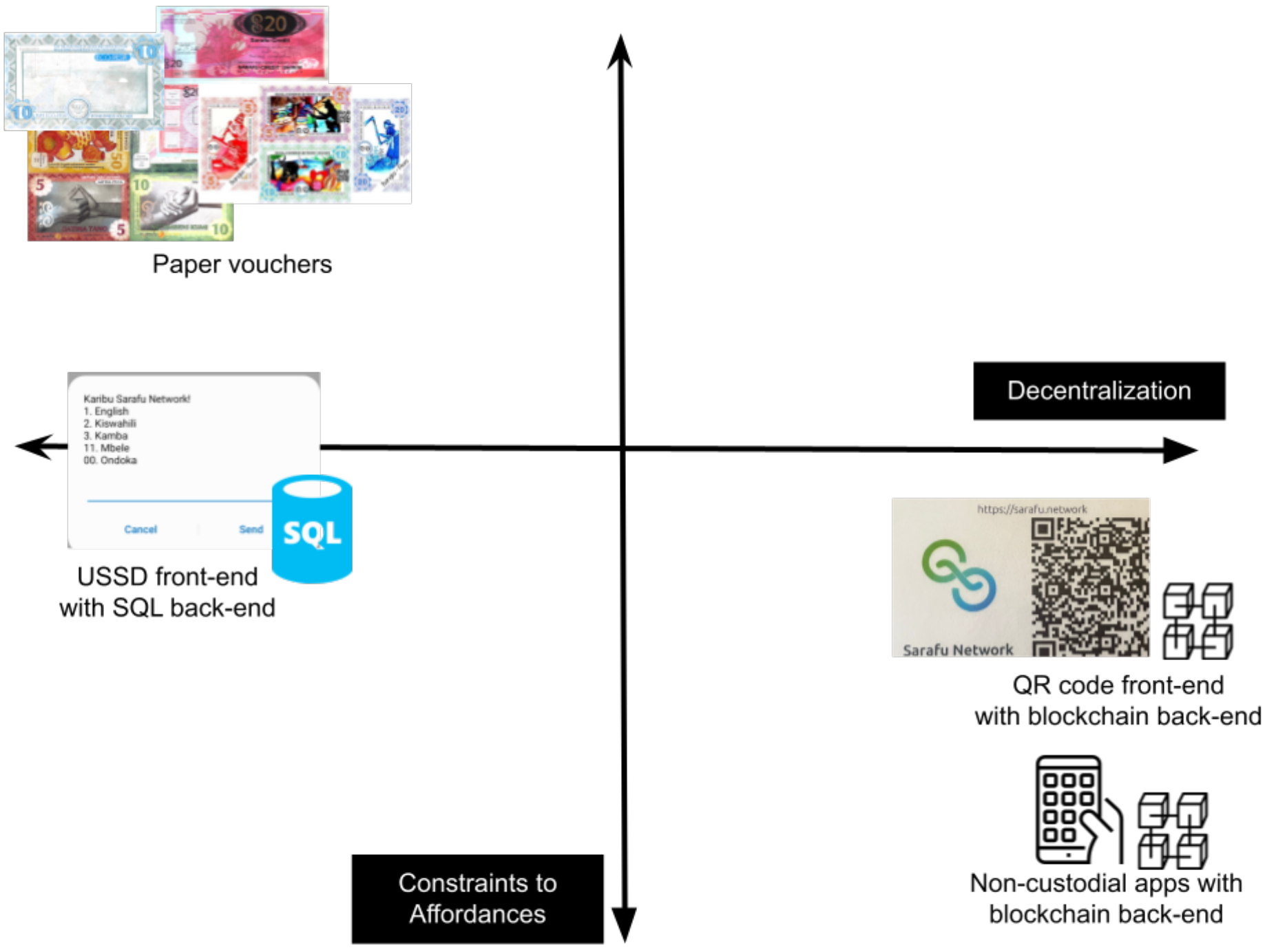

*Source: Author with images from Grassroots Economics*

**NOTES**

---

[1] USSD is a communications protocol used by second-generation (2G) digital cellular networks phones for real-time, session-based interaction with a mobile network, often using short codes like *123#.

[2] SQL is a standardized programming language designed for managing and querying data held in a relational database management system.

[3] The foundation has published their technological documentation at https://docs.grassecon.org/ (last accessed 1 November 2025)

[4] The specific amount of 400 was during the Bangla-Pesa period where every participating business in the BBN would be granted 400 Bangla-Pesa (Patinkin, 2013)

## ACKNOWLEDGEMENTS

The author acknowledges the supervision of Dr. Jérôme Blanc and Dr. Mara Giua throughout this research; the journal reviewers' constructive suggestions; and the assistance of Nick Lunch, Zachary Marlow, Emmanuel Mbui, Jacob Mwatumbi Chaka, Wilfred Mndungu Chibwara, William Ruddick, and Grassroots Economics during fieldwork. All interpretations and remaining errors are the author's own.

## FUNDING

This work is the outcome of the research developed during the Economics for the Global Transition (EPOG+) Erasmus Mundus Joint Master's Degree at Università degli studi Roma Tre, Université de technologie de Compiègne, Sorbonne Université and Université Paris Cité; and was financially supported by the Outgoing Mobility Grant of the SMARTS-UP programme at the Université Paris Cité.

## DECLARATIONS

**Conflicts of interest to declare:** The author declares no conflicts of interest.

**Ethical approval:** This study was conducted in accordance with institutional and professional ethical standards and under a tripartite Internship Agreement between the author, Grassroots Economics, and Université Paris Cité, which specified confidentiality and ethical obligations.

**Informed consent:** Explicit informed consent was obtained for all interviews, focus groups, and workshops; for participants with limited literacy, information was read aloud in Kiswahili and local languages with a translator, including consent for audio recording

**Author contributions:** The author solely designed the study, conducted data collection and analysis, and wrote the manuscript.

**Data availability:** The data generated and analyzed during the current study are not publicly available due to confidentiality and ethical restrictions but are available from the author upon reasonable request.

## ABOUT THE AUTHOR

Patricia Marcella Evite is a Marie Skłodowska-Curie Doctoral Researcher from the University of Naples Federico II, Italy, working at the intersection of finance and artificial intelligence. This article draws on research conducted during her master including fieldwork on community currencies and blockchain technology. Her research interests combine social science, economics, computational methods, and human–computer interaction to examine financial systems.

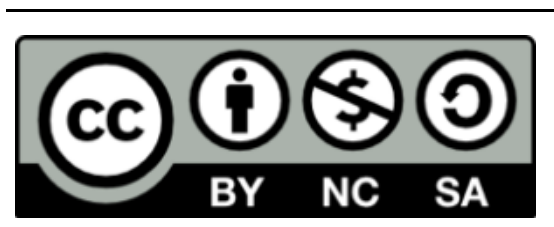